%% file: oscillation_prl.tex
\documentclass[10pt,aps,prl,reprint,twocolumn,amsmath,superscriptaddress,preprintnumbers,floatfix,nofootinbib,showpacs]{revtex4-1}

\usepackage{siunitx}
\usepackage{tikz}
\usepackage{tikz-3dplot}
\usepackage{standalone}
\usepackage{graphicx}
\usepackage{xcolor}
\usepackage{hyperref}
\usepackage{mathtools}
\usepackage{nicefrac}
\usepackage{amssymb}
\usepackage[utf8]{inputenc}
\usepackage[T1]{fontenc}
\usepackage{natbib}
\usepackage{subfigure}

\usepackage{ulem}

\begin{document}

\title{Phase Measurement for Driven Spin Oscillations in a Storage Ring}

\input{authors}

%%%%%%%%%%%%%%%%%%%%%%%%%%%%%%%%%%%%%%%%%%%%%%%%%%%%%%%%%%%%%%%%%%%%%%%%%%%%%%%%%%%%%%%%%%%%%%%%%%%%%%%%%%%%%%%%%%%%%%%%%%%%%%
\begin{abstract}
% Radio frequency devices are an essential tool for manipulating particle spins in a storage ring.
This paper reports the first simultaneous measurement of the horizontal and vertical components of the polarization vector in a storage ring under the influence of a radio frequency (rf) solenoid.
The experiments were performed at the Cooler Synchrotron COSY in Jülich using a vector polarized, bunched $0.97\,\textrm{GeV/c}$ deuteron beam.

Using the new spin feedback system, we set the initial phase difference between the solenoid field and the precession of the polarization vector to a predefined value.
The feedback system was then switched off, allowing the phase difference to change over time, and the solenoid was switched on to rotate the polarization vector.
We observed an oscillation of the vertical polarization component and the phase difference.
The oscillations can be described using an analytical model.

The results of this experiment also apply to other rf devices with horizontal magnetic fields, such as Wien filters.
The precise manipulation of particle spins in storage rings is a prerequisite for measuring the electric dipole moment (EDM) of charged particles.
\end{abstract}

%%%%%%%%%%%%%%%%%%%%%%%%%%%%%%%%%%%%%%%%%%%%%%%%%%%%%%%%%%%%%%%%%%%%%%%%%%%%%%%%%%%%%%%%%%%%%%%%%%%%%%%%%%%%%%%%%%%%%%%%%%%%%%
% \pacs{13.40.Em, 11.30.Er, 29.20.D, 29.20.dg, 29.20.db}
%%%%%%%%%%%%%%%%%%%%%%%%%%%%%%%%%%%%%%%%%%%%%%%%%%%%%%%%%%%%%%%%%%%%%%%%%%%%%%%%%%%%%%%%%%%%%%%%%%%%%%%%%%%%%%%%%%%%%%%%%%%%%%
\maketitle
%%%%%%%%%%%%%%%%%%%%%%%%%%%%%%%%%%%%%%%%%%%%%%%%%%%%%%%%%%%%%%%%%%%%%%%%%%%%%%%%%%%%%%%%%%%%%%%%%%%%%%%%%%%%%%%%%%%%%%%%%%%%%%

Experiments with polarized beams play an important role in accelerator physics, particularly in the search for an electric dipole moment (EDM) of charged elementary particles.
For EDM experiments, the precise measurement and control of the polarization vector is an essential prerequisite.

In this work, we describe the first simultaneous measurement of the horizontal and vertical components of the polarization vector in a particle accelerator under the influence of an rf solenoid.
An analytical model is derived, which is compared to the data.
The work is based on earlier publications; it makes use of the \SI{1000}{\second} spin coherence time at the Cooler Synchrotron (COSY) \cite{Guidoboni:2016bdn}, measurements of the fast (\SI{120}{\kilo\hertz}) precession in the horizontal plane \cite{Eversmann:2015jnk, PhysRevSTAB.17.052803} and a polarization feedback system, which is used to select the initial conditions for the measurements presented here \cite{feedbackPRL}.

Solenoid-induced spin resonances have been studied at COSY before, but the earlier experiments could only measure the vertical, not the horizontal, component of the polarization of the deuteron beam.
The only possible initial state was upward or downward polarization.
Additionally, the analytical model presented here is simpler than the numerical multiplication of rotation matrices used in the earlier publication \cite{SpinResonance2012}.

The presented experiment was performed at COSY under conditions similar to those described in \cite{feedbackPRL}.
A deuteron beam with a vertical vector polarization was injected into COSY and accelerated to a momentum of $970\,\mathrm{MeV/c}$.
The beam was electron-cooled to reduce the emittance.

\begin{figure}
  %  \includestandalone[width=0.7\columnwidth]{fig/anglesStandalone.tex}
    \includegraphics[width=0.7\columnwidth]{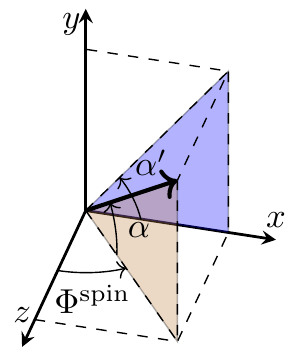}
  \caption{Coordinate system used to describe the spin motion.
  The $z$-axis is defined by the nominal beam momentum, the $y$-axis points upward and the $x$-axis to the side.
  $\alpha$ is the angle between the polarization vector and the horizontal plane.
  $\alpha'$ is the angle between the projection of the polarization vector onto the $xy$-plane and the $x$-axis.
  $\phi^{\textrm{spin}}$ is the phase of the spin rotation.}
  \label{fig:freeOscDiagram}
\end{figure}

Figure \ref{fig:freeOscDiagram} shows the coordinate system used to describe the polarization vector under the influence of the solenoid.
The polarization vector can be described using the precession phase $\phi^{\textrm{spin}}$ and the angle $\alpha$ between the polarization vector and the horizontal plane, where $\tan\alpha = P_{y}/\sqrt{P_{x}^2+P_{z}^2}$.
The magnitude of the polarization is treated as constant.

\begin{figure}
  \includegraphics[width=\columnwidth]{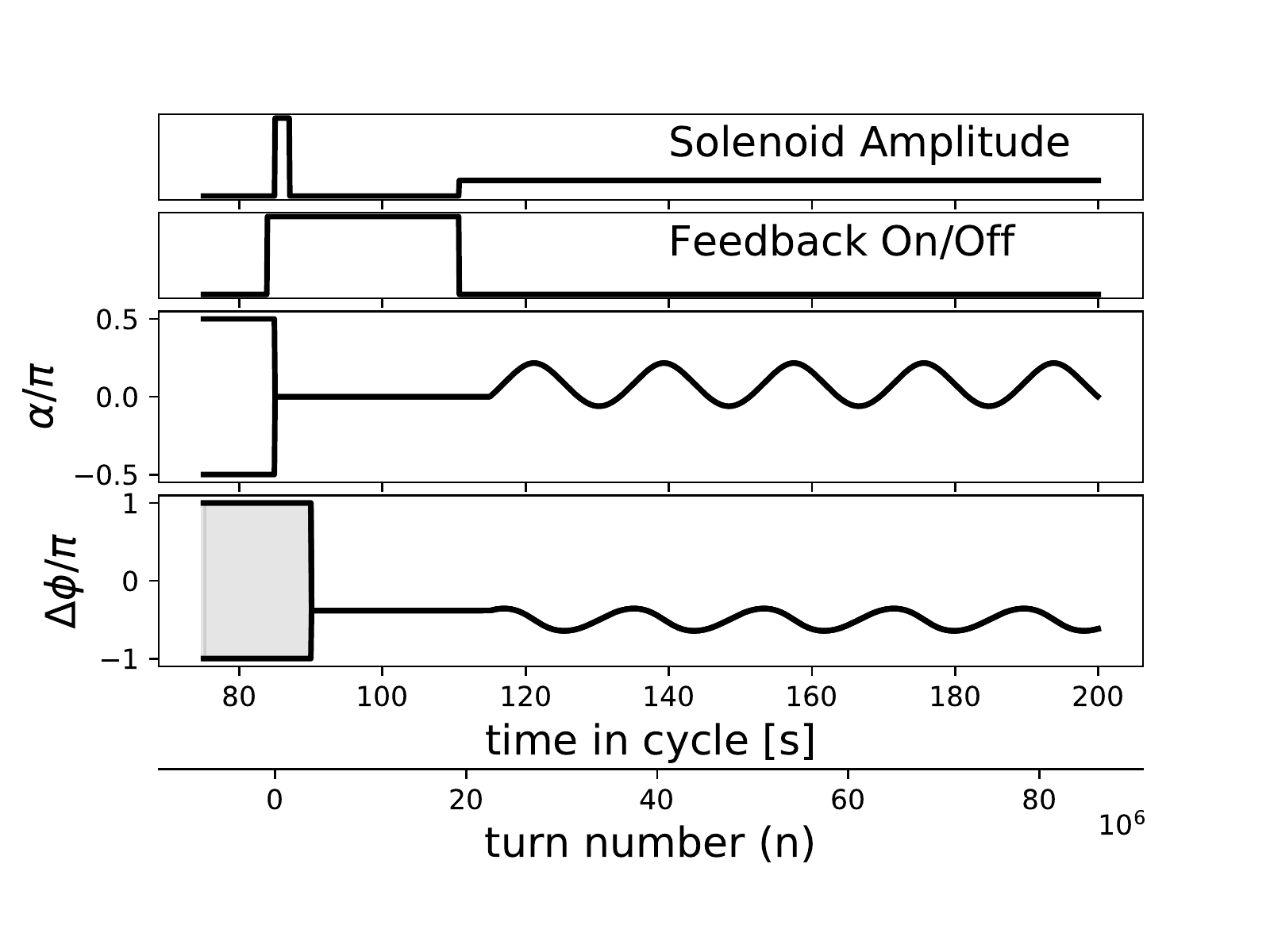}
  \caption{Principle of the experiment.
  $\Delta\phi$ can take any value in the shaded region. $\alpha$ is initially $\pm\pi/2$.
  The first solenoid pulse tilts the polarization into the horizontal plane.
  The feedback system is then switched on to set a certain value of $\Delta\phi$.
  After that, the solenoid is switched on again at a lower amplitude and the feedback is switched off, leading to oscillations.}
  \label{fig:expPrinciple}
\end{figure}

Figure \ref{fig:expPrinciple} shows the basic principle of the experiment.
After \SI{85}{\second}, the polarization was rotated into the horizontal plane using the solenoid, and the feedback system was switched on.
This point is defined as turn number $0$.
The feedback system then set the relative phase between spin precession and the solenoid rf voltage to a predefined value $\Delta\phi_0$.
\SI{115}{\second} after the start of the cycle, at turn number $22.9\cdot10^6$, the solenoid was switched on again while the feedback system was switched off, allowing the spins to precess independently from the solenoid.
This is the important difference from the procedure described in \cite{feedbackPRL}, where the feedback system remained active for the whole cycle.
The experiment was repeated for 16 values of $\Delta\phi_0$ between $-\pi$ and $\pi$.

It was observed that both the vertical polarization and the relative phase oscillate with the same frequency in the order of \SI{0.1}{\hertz}, proportional to the solenoid amplitude.
This behavior can be explained by an analytical model depending on four parameters, which can be chosen as the solenoid amplitude, the difference of the solenoid frequency from the ideal resonance, the initial relative phase, and the initial angle $\alpha$ between the polarization vector and the horizontal plane.
This analytic description is equivalent to the Single Resonance Model (SRM) decribed in~\cite{Mane:2005xh}.

The magnetic fields in the storage ring cause a precession about the $y$-axis of $2\pi \nu_{s}$ per turn, where $\nu_{s} = f^{\textrm{spin}}/f^{\textrm{COSY}} \approx \gamma G \approx -0.16$ is the spin tune.
The solenoid rotates the polarization about the $z$-axis by an angle $k\cdot\sin\phi^{\textrm{sol}}$, where $k$ is proportional to the amplitude of the solenoid rf signal and $\phi^{\textrm{sol}}$ is its phase.
As long as $\phi^{\textrm{spin}}$ is not $\pm\pi/2$, a rotation about the $z$-axis also affects $\phi^{\textrm{spin}}$.
This way, the solenoid can advance or delay the precession in the horizontal plane.

From turn number $n$ to $n+1$, the change in $\alpha$, $\phi^{\textrm{spin}}$ and the solenoid phase $\phi^{\textrm{sol}}$ is
\begin{equation}
  \begin{aligned}
    \phi^{\textrm{spin}}_{n+1} &= \phi^{\textrm{spin}}_{n} + 2\pi\nu_{s} + k \sin \phi^{\textrm{sol}}_{n} \left.\frac{d \phi^{\textrm{spin}}}{d\alpha'}\right|_{P_{z}=\text{const}} \\
    \alpha_{n+1} &= \alpha_{n} + k \sin \phi^{\textrm{sol}}_{n} \left.\frac{d \alpha}{d\alpha'}\right|_{P_{z}=\text{const}} \\
    \phi^{\textrm{sol}}_{n+1} &= \phi^{\textrm{sol}}_{n} + 2\pi\nu_{s}^{\textrm{sol}},
    \label{eq:turnByTurnChange}
  \end{aligned}
\end{equation}
where $\nu_{s}^{\textrm{sol}}=f^{\textrm{sol}}/f^{\textrm{COSY}}$ is the number of oscillations the solenoid performs per turn, which is equal to $\nu_{s}$ plus an integer number when the solenoid is on resonance.
It will become apparent later that distinguishing between $\nu_{s}$ and $\nu_{s}^{\textrm{sol}}$ is essential to describe the data.
The definition of $\alpha'$ is indicated in Figure \ref{fig:freeOscDiagram}.

Eqs. (\ref{eq:turnByTurnChange}) can be simplified by substituting $\Delta\phi_{n} = \phi_{n}^{\textrm{spin}} - \phi_{n}^{\textrm{sol}} = \phi^{\textrm{spin}} - 2\pi n \nu_{s}^{\textrm{sol}}$ and the geometrical derivatives $\left.\frac{d \phi^{\textrm{spin}}}{d\alpha'}\right|_{P_{z}=\text{const}} = -\tan \alpha \cos\phi^{\textrm{spin}}$ and $\left.\frac{d \alpha}{d\alpha'}\right|_{P_{z}=\text{const}} = \sin \phi^{\textrm{spin}}$.
This results in:
\begin{equation}
  \begin{aligned}
    \Delta\phi_{n+1} &= \Delta\phi_{n} + 2\pi (\nu_{s}-\nu_{s}^{\textrm{sol}}) \\
    &- k\sin(2\pi n \nu_{s}^{\textrm{sol}}) \tan \alpha_{n} \cos (\Delta\phi_{n} + 2\pi n \nu_{s}^{\textrm{sol}}) \\
    \alpha_{n+1} &= \alpha_{n} + k\sin(2\pi n \nu_{s}^{\textrm{sol}}) \sin (\Delta\phi_{n} + 2\pi n \nu_{s}^{\textrm{sol}}). \\
  \end{aligned}
\end{equation}
The initial phase of the solenoid can be set to zero without loss of generality.
The oscillation has a rapid component with a frequency proportional to $\nu_{s} f^{\textrm{COSY}}\approx 120\,\mathrm{kHz}$ and a slow component proportional to $k f^{\textrm{COSY}}$.
Since we are interested in the polarization evolution over a longer time scale, we replace the fast component with its time average, which yields two coupled differential equations:
\begin{equation}
  \begin{aligned}
    \frac{d\alpha}{dn} &= \frac{k}{2} \cos \Delta\phi, \\
    \frac{d\Delta\phi}{dn} &= \frac{k}{2} \left(\tan\alpha \sin \Delta\phi + q \right).
  \end{aligned}
  \label{eq:offResSolEquations}
\end{equation}
The parameter $q=4\pi(\nu_{s} - \nu_{s}^{\textrm{sol}})/k$ indicates how close the solenoid is to spin resonance.
$q$ was typically in the order of 1, with $k\approx10^{-6}$.

The solution to Eqs. (\ref{eq:offResSolEquations}) can be written as:
\begin{equation}
  \begin{aligned}
    \sin\alpha(n) &= A_{1} \sin\left(A_{2}+n A_{3}\right) - A_{4}\\
    \cos\Delta\phi(n) &= \frac{A_{1} \sqrt{1+q^2} \cos\left(A_{2}+ n A_{3}\right)}{\sqrt{1-\left(A_{1}\sin\left(A_{2}+ n A_{3}\right)-A_{4}\right)^2}}\\
    \sin\Delta\phi(n) &= \frac{C+q\sin\alpha}{\cos\alpha},
  \end{aligned}
  \label{eq:offResSolSolutions}
\end{equation}
with the parameters
\begin{equation}
  \begin{aligned}
    A_{1} &= \frac{\sqrt{1+q^2-C^2}}{1+q^2}\\
    A_{2} &= \begin{dcases}
      \arcsin\left(\frac{\sin\alpha_0+A_4}{A_1}\right) & |\Delta\phi_{0}| < \pi/2 \\
      \pi - \arcsin\left(\frac{\sin\alpha_0+A_4}{A_1}\right) & |\Delta\phi_{0}| > \pi/2 \\
    \end{dcases}\\
    A_{3} &= \frac{k}{2} \sqrt{1+q^2}\\
    A_{4} &= \frac{Cq}{1+q^2}.
  \end{aligned}
  \label{eq:offResSolParams}
\end{equation}
$A_1$, $A_4$ and $q$ are not independent but coupled via the following equation:
\begin{equation}
  \label{eq:1plusqSolution}
  1+q^2 = \frac{1+A_1^2-A_4^2}{2A_1^2} \pm \sqrt{\left(\frac{1+A_1^2-A_4^2}{2A_1^2}\right)^2-\frac{1}{A_1^2}}.
\end{equation}
The positive sign in (\ref{eq:1plusqSolution}) corresponds to unbound solutions, the negative sign to bound ones (defined below).
The quantity $C=\cos\alpha\sin\Delta\phi-q\sin\alpha$ is conserved in equations (\ref{eq:offResSolEquations}).
The parameters $\alpha_0=0$ (which simplifies $A_{2}$ in Eq. (\ref{eq:offResSolParams})) and $\Delta\phi_0$ are the initial values.
By setting $\Delta\phi$ to a certain value at the beginning of a measurement, it is possible to set the amplitude of the oscillations.
The farther $\Delta\phi_0$ is from $\pm\pi/2$, the larger the amplitudes of the oscillations in $\alpha$ and $\Delta\phi$.

\begin{figure}
  \includegraphics[width=\columnwidth]{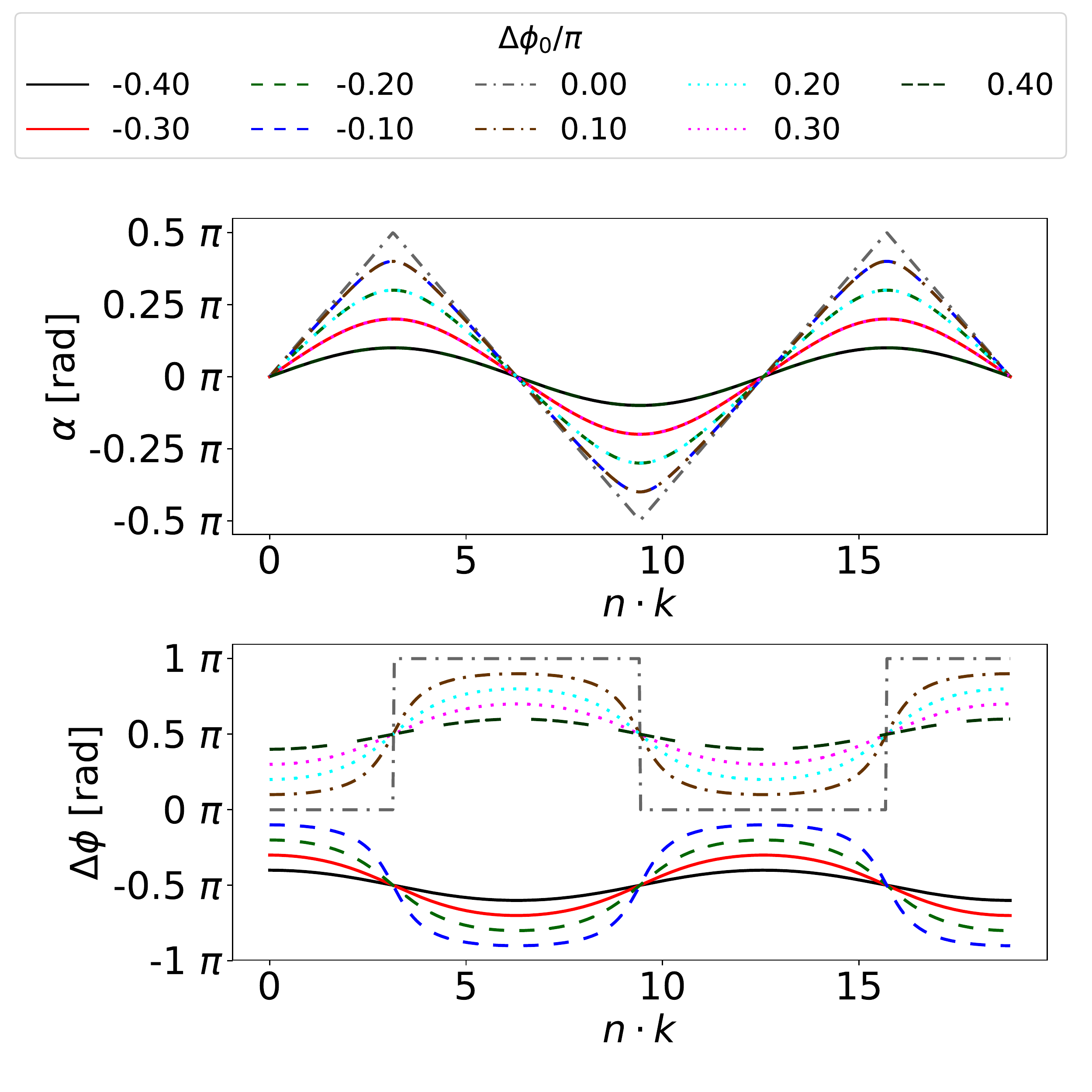}
  \caption{Analytical solutions (Eq. (\ref{eq:offResSolSolutions})) for the on-resonance case $q=0$ and $\alpha_{0}=0$ as a function of $nk$ for different values of $\Delta\phi_{0}$.}
  \label{fig:onResSolution}
\end{figure}
\begin{figure}
  \includegraphics[width=\columnwidth]{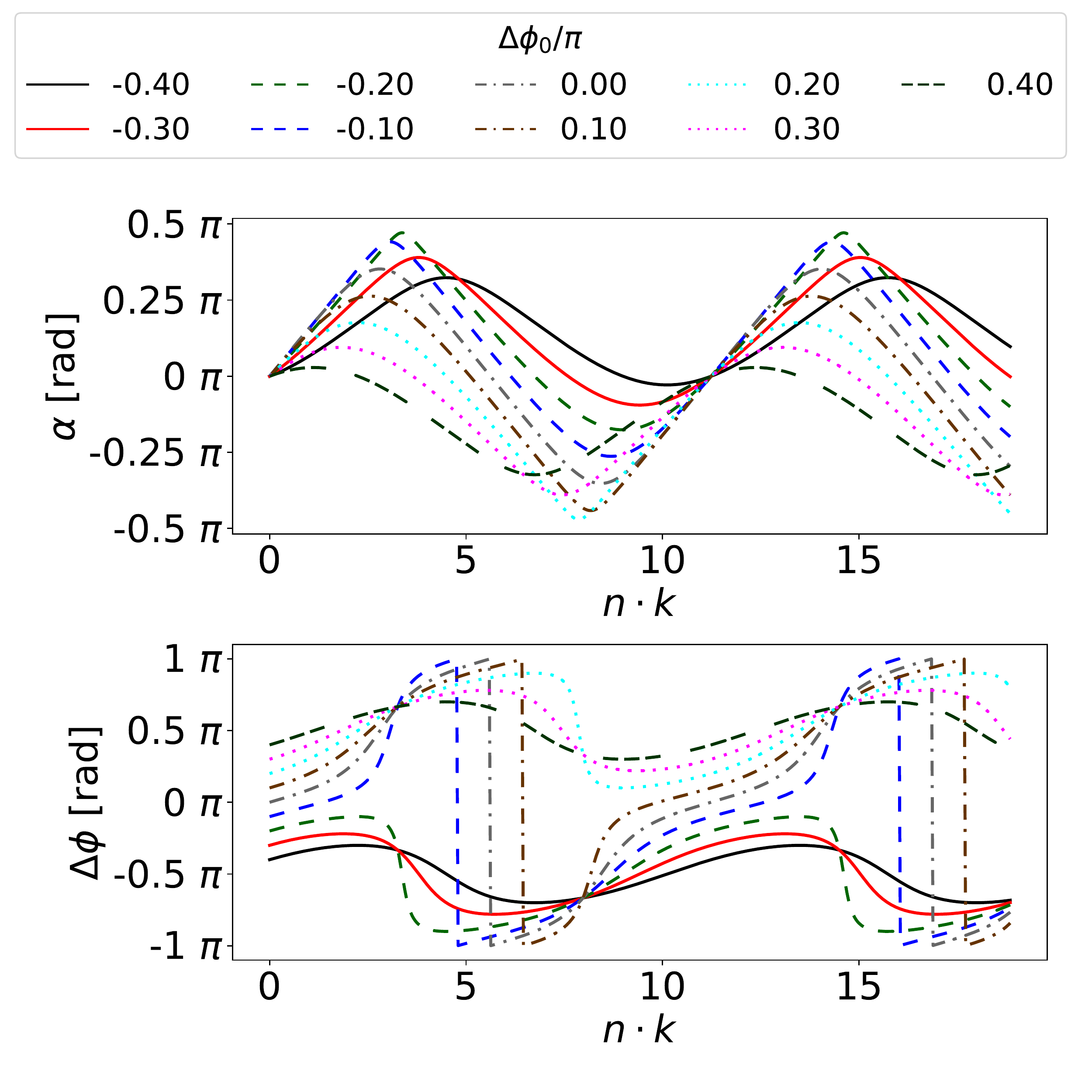}
  \caption{Analytical solutions (Eq. (\ref{eq:offResSolSolutions})) for the off-resonance case $q=0.5$. The other parameters are the same as in Figure \ref{fig:onResSolution}.}
  \label{fig:offResSolution}
\end{figure}
Figure \ref{fig:onResSolution} shows some solutions for the ideal on-resonance case ($q=0$).
For low amplitudes the oscillations approach a sinusoidal shape.
As the amplitude increases, the $\alpha$ curve approaches a triangular function and the $\Delta\phi$ curve approaches a step function.

Figure \ref{fig:offResSolution} shows solutions for the off-resonance case at $q=0.5$.
For every solution, another one can be obtained by the transformation $\Delta\phi \rightarrow \Delta\phi+\pi$ and $\alpha \rightarrow -\alpha$.
These solutions are omitted in the plots for clarity.
Solutions can be divided into two classes: bound solutions, in which $\Delta\phi$ oscillates around $\pm\pi/2$ without crossing zero, and unbound solutions, in which $\Delta\phi$ moves over the whole range from $-\pi$ to $\pi$.
The latter only occur in the off-resonance case.
Solutions are unbound if and only if $|C|<|q|$.
For $q=0$, the amplitudes of the oscillations in $\alpha$ and $\Delta\phi$ are equal (see also \cite{lee1997spin}).

An off-resonant solenoid frequency has two more effects.
% First, the angles no longer oscillate around $\Delta\phi = \pm \pi/2$ and $\alpha=0$ but around $\Delta\phi = \pm \pi/2$ and $\alpha=\mp \arctan q$.
First, $\alpha$ no longer oscillates around $0$ but around $\pm \arctan q$, where the plus sign applies to cases in which $\Delta\phi$ oscillates around $-\pi/2$ and the minus sign applies to cases in which $\Delta\phi$ oscillates around $\pi/2$.
Secondly, the frequency of the oscillation increases by a factor of $\sqrt{1+q^2}$.

\begin{figure}
  \includegraphics[width=\columnwidth]{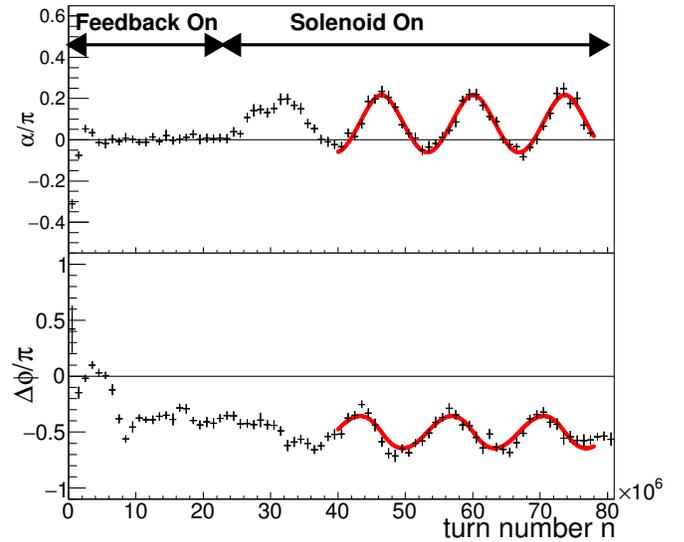}
  \caption{Observation of bound oscillation of $\Delta\phi$ around $-\pi/2$. The fit results correspond to $k=(8.95\pm0.11)\cdot10^{-7}$, $q= 0.250\pm0.027$ and $C=-0.934\pm0.013$ at a $\chi^2/\textrm{NDF}$ of $101.4/74$.}
  \label{fig:onResData}
\end{figure}
\begin{figure}
  \includegraphics[width=\columnwidth]{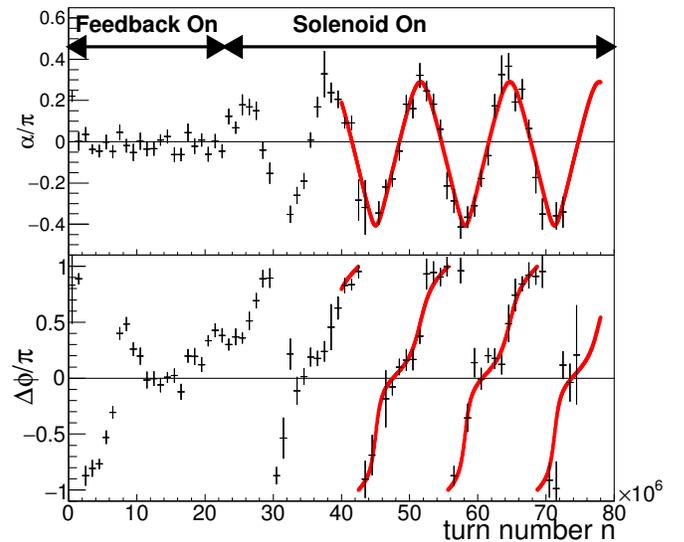}
  \caption{Observation of unbound oscillation of $\Delta\phi$. The fit results correspond to $k=(8.51\pm0.58)\cdot10^{-7}$, $q= 0.52\pm0.16$ and $C=0.205\pm0.087$ at a $\chi^2/\textrm{NDF}$ of $82.8/74$.}
  \label{fig:offResData}
\end{figure}
Figs. \ref{fig:onResData} and \ref{fig:offResData} show measured values of $\Delta\phi$ and $\alpha$ along with fits of Eq. (\ref{eq:offResSolSolutions}) for different starting values $\Delta\phi_{0}$.
The settings for the solenoid were kept constant.
The fits were performed using a combined $\chi^2$-minimization for the $\alpha$ and $\Delta\phi$ data.
In Figure \ref{fig:onResData}, $\Delta \phi$ oscillates around $-\pi/2$, while $\alpha$ oscillates around a value significantly greater than zero.
In Figure \ref{fig:offResData}, $\Delta \phi$ is not confined to a limited interval.
Runs for all values of $\Delta\phi_0$ are described well by the model, confirming its validity.

Another interesting property of the equations of motion is that they simplify to a circular motion at a constant speed if $\alpha$ and $\Delta\phi$ are interpreted as the elevation and azimuth angles of a spherical coordinate system (see also \cite{saleevSpinTuneMapping})
\begin{equation}
  \begin{aligned}
    x_1 &= \cos\alpha \cos\Delta\phi \\
    x_2 &= \cos\alpha \sin\Delta\phi \\
    x_3 &= \sin\alpha.
    \label{eq:cartesianSystem}
  \end{aligned}
\end{equation}
For an on-resonance solenoid, the circles lie in the $xz$-plane.
In the off-resonance case, they are tilted by $\arctan q$.
\begin{figure}
      \includegraphics[width=0.95\columnwidth]{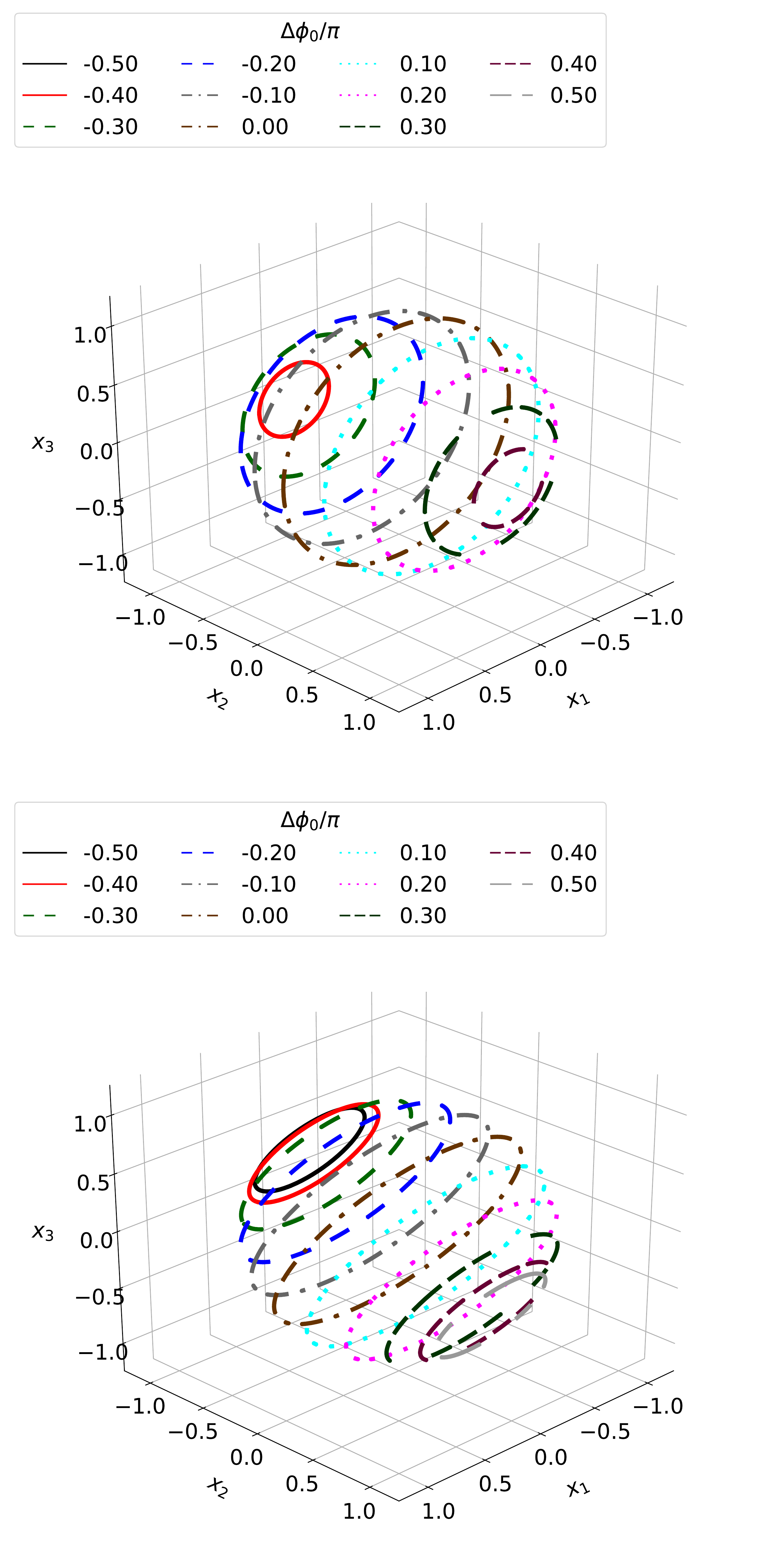}\\
  \caption{Representation of on-resonance ($q=0$, top) and off-resonance ($q=0.5$, bottom) solutions in the coordinate system defined in Eq. (\ref{eq:cartesianSystem})}
  \label{fig:3dSolutions}
\end{figure}

\begin{figure}
  \includegraphics[width=\columnwidth]{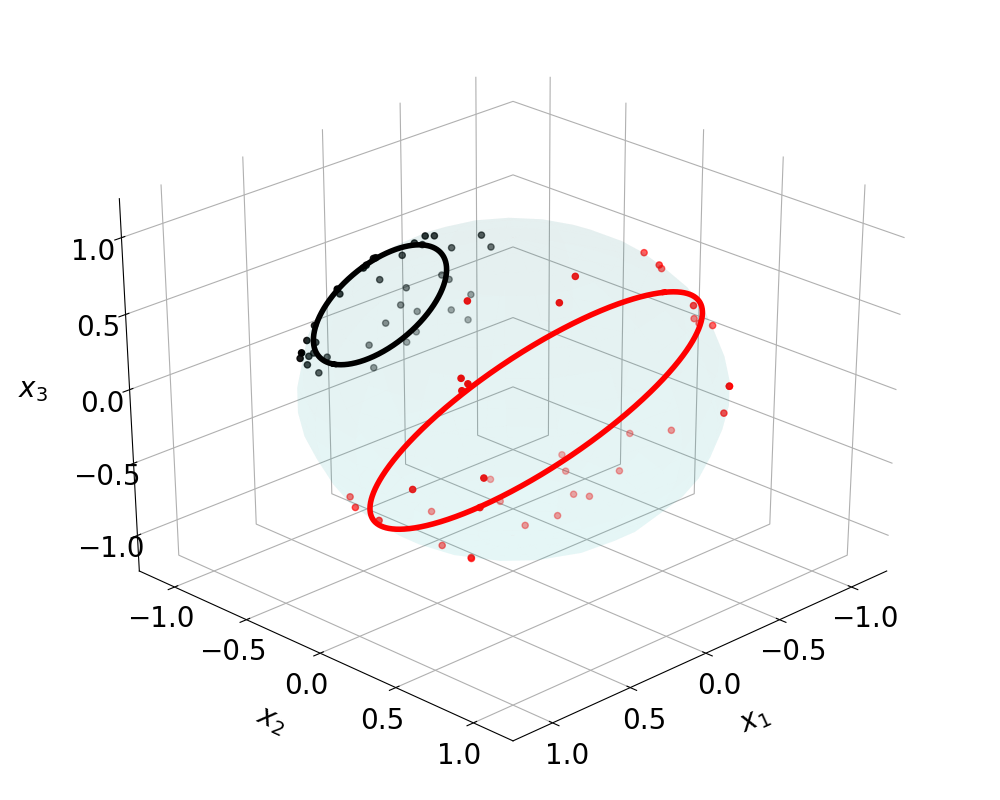}
  \caption{Representation of the data and fits from Figs. \ref{fig:onResData} (black) and \ref{fig:offResData} (red) in the Cartesian coordinate system defined in Eq. (\ref{eq:cartesianSystem}).
  All points and curves lie on the unit sphere.}
  \label{fig:3dData}
\end{figure}
Figure \ref{fig:3dSolutions} shows representative solutions of the equations of motion for the on-resonance and off-resonance cases.
The tilt of the circles in the off-resonance case is clearly visible.
Figure \ref{fig:3dData} shows the results of the fit examples under this transformation.

Manipulating the particles using rf devices is important for a future measurement of electric dipole moments using the so-called Wien Filter method \cite{Rathmann:2013rqa, PhysRevSTAB.16.114001, Slim:2016pim}.
A Wien Filter is a device using electric and magnetic fields that are orthogonal to each other and to the beam, so that the Lorentz force is zero at the design momentum.
If the magnetic field is in the ring plane, a Wien Filter can cause an oscillation in the vertical polarization, which is similar to the one observed here, and which follows the same basic equations \cite{mey2015WienFilter}.
A related model for an rf Wien Filter is presented in \cite{saleevSpinTuneMapping}.

In conclusion, we have measured the effect of an rf solenoid on the vector polarization of deuterons in a particle accelerator, and described it using an analytical model.
The phase between the spin precession and the solenoid frequency was measured for the first time.
The new polarization feedback system was used to set the initial conditions.
Spin manipulation using rf devices will be important in later experiments to measure electric dipole moments in a storage ring using the Wien filter method.

\begin{acknowledgments}
The authors wish to thank the staff of COSY for providing excellent working conditions and for their support concerning the technical aspects of this experiment.
This work has been financially supported by Forschungszentrum J\"ulich GmbH, Germany, via COSY FFE, by an ERC Advanced-Grant (srEDM \# 694390) of the European Union, the European Union Seventh Framework Programme (FP7/2007-2013) under Grant Agreement No. 283286, by the Shota Rustaveli National Science Foundation of Georgia (SRNSF grant No DI/13/6-200/14), by a grant from the Russian Science Foundation (Grant No. RNF-16-12-10151) and by IBS-R017-D1 of the Republic of Korea.
\end{acknowledgments}

\bibliographystyle{apsrev4-1}
\bibliography{literature_edm}
\end{document}

%% file: authors.tex
\author{N.~Hempelmann}
\affiliation{III. Physikalisches Institut B, RWTH Aachen University, 52056 Aachen, Germany}
\affiliation{Institut f\"ur Kernphysik, Forschungszentrum J\"ulich, 52425 J\"ulich, Germany}
\author{V.~Hejny}
\affiliation{Institut f\"ur Kernphysik, Forschungszentrum J\"ulich, 52425 J\"ulich, Germany}
\author{J.~Pretz}
\affiliation{III. Physikalisches Institut B, RWTH Aachen University, 52056 Aachen, Germany}
\affiliation{Institut f\"ur Kernphysik, Forschungszentrum J\"ulich, 52425 J\"ulich, Germany}
\affiliation{JARA--FAME (Forces and Matter Experiments), Forschungszentrum J\"ulich and RWTH Aachen University, Germany}
\author{H.~Soltner}
\affiliation{Zentralinstitut f\"ur Engineering, Elektronik und Analytik (ZEA-1), Forschungszentrum J\"ulich, 52425 J\"ulich, Germany}
\author{W.~Augustyniak}
\affiliation{Department of Nuclear Physics, National Centre for Nuclear Research, 00681 Warsaw, Poland}
\author{Z.~Bagdasarian}
\affiliation{High Energy Physics Institute, Tbilisi State University, 0186 Tbilisi, Georgia}
\affiliation{Institut f\"ur Kernphysik, Forschungszentrum J\"ulich, 52425 J\"ulich, Germany}
\author{M.~Bai}
\affiliation{Institut f\"ur Kernphysik, Forschungszentrum J\"ulich, 52425 J\"ulich, Germany}
\affiliation{JARA--FAME (Forces and Matter Experiments), Forschungszentrum J\"ulich and RWTH Aachen University, Germany}
\author{L.~Barion}
\affiliation{University of Ferrara and INFN, 44100 Ferrara, Italy}
\author{M.~Berz}
\affiliation{Department of Physics and Astronomy, Michigan State University,  East Lansing, Michigan 48824, USA}
\author{S.~Chekmenev}
\affiliation{III. Physikalisches Institut B, RWTH Aachen University, 52056 Aachen, Germany}
\author{G.~Ciullo}
\affiliation{University of Ferrara and INFN, 44100 Ferrara, Italy}
\author{S.~Dymov}
\affiliation{Institut f\"ur Kernphysik, Forschungszentrum J\"ulich, 52425 J\"ulich, Germany}
\affiliation{Laboratory of Nuclear Problems, Joint Institute for Nuclear Research, 141980 Dubna, Russia}
\author{D. Eversmann}
\affiliation{III. Physikalisches Institut B, RWTH Aachen University, 52056 Aachen, Germany}
\author{M.~Gaisser}
\affiliation{Center for Axion and Precision Physics Research, Institute for Basic Science, Daejeon 34051, Republic of Korea}
\affiliation{III. Physikalisches Institut B, RWTH Aachen University, 52056 Aachen, Germany}
\author{R.~Gebel}
\affiliation{Institut f\"ur Kernphysik, Forschungszentrum J\"ulich, 52425 J\"ulich, Germany}
\author{K.~Grigoryev}
\affiliation{III. Physikalisches Institut B, RWTH Aachen University, 52056 Aachen, Germany}
\author{D.~Grzonka}
\affiliation{Institut f\"ur Kernphysik, Forschungszentrum J\"ulich, 52425 J\"ulich, Germany}
\author{G.~Guidoboni}
\affiliation{University of Ferrara and INFN, 44100 Ferrara, Italy}
\author{D.~Heberling}
\affiliation{Institut f\"ur Hochfrequenztechnik, RWTH Aachen University, 52056 Aachen, Germany}
\affiliation{JARA--FAME (Forces and Matter Experiments), Forschungszentrum J\"ulich and RWTH Aachen University, Germany}
\author{J.~Hetzel}
\affiliation{Institut f\"ur Kernphysik, Forschungszentrum J\"ulich, 52425 J\"ulich, Germany}
\author{F.~Hinder}
\affiliation{III. Physikalisches Institut B, RWTH Aachen University, 52056 Aachen, Germany}
\affiliation{Institut f\"ur Kernphysik, Forschungszentrum J\"ulich, 52425 J\"ulich, Germany}
\author{A.~Kacharava}
\affiliation{Institut f\"ur Kernphysik, Forschungszentrum J\"ulich, 52425 J\"ulich, Germany}
\author{V.~Kamerdzhiev}
\affiliation{Institut f\"ur Kernphysik, Forschungszentrum J\"ulich, 52425 J\"ulich, Germany}
\author{I.~Keshelashvili}
\affiliation{Institut f\"ur Kernphysik, Forschungszentrum J\"ulich, 52425 J\"ulich, Germany}
\author{I.~Koop}
\affiliation{Budker Institute of Nuclear Physics, 630090 Novosibirsk, Russia}
\author{A.~Kulikov}
\affiliation{Laboratory of Nuclear Problems, Joint Institute for Nuclear Research, 141980 Dubna, Russia}
\author{A.~Lehrach}
\affiliation{Institut f\"ur Kernphysik, Forschungszentrum J\"ulich, 52425 J\"ulich, Germany}
\affiliation{JARA--FAME (Forces and Matter Experiments), Forschungszentrum J\"ulich and RWTH Aachen University, Germany}
\author{P.~Lenisa}
\affiliation{University of Ferrara and INFN, 44100 Ferrara, Italy}
\author{N.~Lomidze}
\affiliation{High Energy Physics Institute, Tbilisi State University, 0186 Tbilisi, Georgia}
\author{B.~Lorentz}
\affiliation{Institut f\"ur Kernphysik, Forschungszentrum J\"ulich, 52425 J\"ulich, Germany}
\author{P.~Maanen}
\affiliation{III. Physikalisches Institut B, RWTH Aachen University, 52056 Aachen, Germany}
\author{G.~Macharashvili}
\affiliation{High Energy Physics Institute, Tbilisi State University, 0186 Tbilisi, Georgia}
\affiliation{Laboratory of Nuclear Problems, Joint Institute for Nuclear Research, 141980 Dubna, Russia}
\author{A.~Magiera}
\affiliation{Institute of Physics, Jagiellonian University, 30348 Cracow, Poland}
\author{D.~Mchedlishvili}
\affiliation{High Energy Physics Institute, Tbilisi State University, 0186 Tbilisi, Georgia}
\affiliation{Institut f\"ur Kernphysik, Forschungszentrum J\"ulich, 52425 J\"ulich, Germany}
\author{S.~Mey}
\affiliation{III. Physikalisches Institut B, RWTH Aachen University, 52056 Aachen, Germany}
\affiliation{Institut f\"ur Kernphysik, Forschungszentrum J\"ulich, 52425 J\"ulich, Germany}
\author{F.~M\"uller}
\affiliation{III. Physikalisches Institut B, RWTH Aachen University, 52056 Aachen, Germany}
\affiliation{Institut f\"ur Kernphysik, Forschungszentrum J\"ulich, 52425 J\"ulich, Germany}
\author{A.~Nass}
\affiliation{Institut f\"ur Kernphysik, Forschungszentrum J\"ulich, 52425 J\"ulich, Germany}
\author{N.N. Nikolaev}
\affiliation{L.D. Landau Institute for Theoretical Physics, 142432 Chernogolovka, Russia}
\affiliation{Moscow Institute for Physics and Technology, 141700 Dolgoprudny, Russia}
\author{M.~Nioradze}
\affiliation{High Energy Physics Institute, Tbilisi State University, 0186 Tbilisi, Georgia}
\author{A.~Pesce}
\affiliation{University of Ferrara and INFN, 44100 Ferrara, Italy}
\author{D.~Prasuhn}
\affiliation{Institut f\"ur Kernphysik, Forschungszentrum J\"ulich, 52425 J\"ulich, Germany}
\author{F.~Rathmann}
\affiliation{Institut f\"ur Kernphysik, Forschungszentrum J\"ulich, 52425
  J\"ulich, Germany}
\author{M.~Rosenthal}
\affiliation{III. Physikalisches Institut B, RWTH Aachen University, 52056 Aachen, Germany}
\affiliation{Institut f\"ur Kernphysik, Forschungszentrum J\"ulich, 52425 J\"ulich, Germany}
\author{A.~Saleev}
\affiliation{Institut f\"ur Kernphysik, Forschungszentrum J\"ulich, 52425 J\"ulich, Germany}
\affiliation{Samara National Research University, 443086 Samara, Russia}
\author{V.~Schmidt}
\affiliation{III. Physikalisches Institut B, RWTH Aachen University, 52056 Aachen, Germany}
\affiliation{Institut f\"ur Kernphysik, Forschungszentrum J\"ulich, 52425 J\"ulich, Germany}
\author{Y.~Semertzidis}
\affiliation{Center for Axion and Precision Physics Research, Institute for Basic Science, Daejeon 34051, Republic of Korea}
\affiliation{Department of Physics, KAIST, Daejeon 34141, Republic of Korea}
\author{Y.~Senichev}
\affiliation{Institut f\"ur Kernphysik, Forschungszentrum J\"ulich, 52425 J\"ulich, Germany}
\author{V.~Shmakova}
\affiliation{Laboratory of Nuclear Problems, Joint Institute for Nuclear Research, 141980 Dubna, Russia}
\author{A.~Silenko}
\affiliation{Research Institute for Nuclear Problems, Belarusian State University, 220030 Minsk, Belarus}
\affiliation{Bogoliubov Laboratory of Theoretical Physics, Joint Institute for Nuclear Research, 141980 Dubna, Russia}
\author{J.~Slim}
\affiliation{Institut f\"ur Hochfrequenztechnik, RWTH Aachen University, 52056 Aachen, Germany}
\author{A.~Stahl}
\affiliation{III. Physikalisches Institut B, RWTH Aachen University, 52056 Aachen, Germany}%
\affiliation{JARA--FAME (Forces and Matter Experiments), Forschungszentrum J\"ulich and RWTH Aachen University, Germany}
\author{R.~Stassen}
\affiliation{Institut f\"ur Kernphysik, Forschungszentrum J\"ulich, 52425 J\"ulich, Germany}
\author{E.~Stephenson}
\affiliation{Indiana University Center for Spacetime Symmetries, Bloomington,  Indiana 47405, USA}
\author{H.~Stockhorst}
\affiliation{Institut f\"ur Kernphysik, Forschungszentrum J\"ulich, 52425 J\"ulich, Germany}
\author{H.~Str\"oher}
\affiliation{Institut f\"ur Kernphysik, Forschungszentrum J\"ulich, 52425 J\"ulich, Germany}
\affiliation{JARA--FAME (Forces and Matter Experiments), Forschungszentrum J\"ulich and RWTH Aachen University, Germany}
\author{M.~Tabidze}
\affiliation{High Energy Physics Institute, Tbilisi State University, 0186 Tbilisi, Georgia}
\author{G.~Tagliente}
\affiliation{INFN, 70125 Bari, Italy}
\author{R.~Talman}
\affiliation{Cornell University, Ithaca,  New York 14850, USA}
\author{P.~Th\"orngren Engblom}
\affiliation{Department of Physics, KTH Royal Institute of Technology, SE-10691 Stockholm, Sweden}
%\affiliation{University of Ferrara and INFN, 44100 Ferrara, Italy}
%
\author{F.~Trinkel}
\affiliation{III. Physikalisches Institut B, RWTH Aachen University, 52056 Aachen, Germany}
\affiliation{Institut f\"ur Kernphysik, Forschungszentrum J\"ulich, 52425 J\"ulich, Germany}
\author{Yu.~Uzikov}
\affiliation{Laboratory of Nuclear Problems, Joint Institute for Nuclear Research, 141980 Dubna, Russia}
\affiliation{Dubna State University, 141980 Dubna, Russia}
\author{Yu.~Valdau}
\affiliation{Helmholtz-Institut f\"ur Strahlen- und Kernphysik, Universit\"at Bonn, 53115 Bonn, Germany}
\affiliation{Petersburg Nuclear Physics Institute, 188300 Gatchina, Russia}
\author{E.~Valetov}
\affiliation{Department of Physics and Astronomy, Michigan State University,  East Lansing, Michigan 48824, USA}
\author{A.~Vassiliev}
\affiliation{Petersburg Nuclear Physics Institute, 188300 Gatchina, Russia}
\author{C.~Weidemann}
\affiliation{Institut f\"ur Kernphysik, Forschungszentrum J\"ulich, 52425 J\"ulich, Germany}
\author{A.~Wro\'{n}ska}
\affiliation{Institute of Physics, Jagiellonian University, 30348 Cracow, Poland}
\author{P.~W\"ustner}
\affiliation{Zentralinstitut f\"ur Engineering, Elektronik und Analytik (ZEA-2), Forschungszentrum J\"ulich, 52425 J\"ulich, Germany}
\author{P.~Zupra\'nski}
\affiliation{Department of Nuclear Physics, National Centre for Nuclear Research, 00681 Warsaw, Poland}
\author{M.~\.{Z}urek}
\affiliation{Institut f\"ur Kernphysik, Forschungszentrum J\"ulich, 52425 J\"ulich, Germany}
\collaboration{JEDI collaboration}